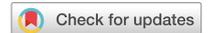

# SCIENTIFIC REPORTS

natureresearch

OPEN

# Multi-AI competing and winning against humans in iterated Rock-Paper-Scissors game

Lei Wang[1], Wenbin Huang[1,2], Yuanpeng Li[1,2], Julian Evans[1] & Sailing He[1,2,3 ✉]

Predicting and modeling human behavior and finding trends within human decision-making processes is a major problem of social science. Rock Paper Scissors (RPS) is the fundamental strategic question in many game theory problems and real-world competitions. Finding the right approach to beat a particular human opponent is challenging. Here we use an AI (artificial intelligence) algorithm based on Markov Models of one fixed memory length (abbreviated as "single AI") to compete against humans in an iterated RPS game. We model and predict human competition behavior by combining many Markov Models with different fixed memory lengths (abbreviated as "multi-AI"), and develop an architecture of multi-AI with changeable parameters to adapt to different competition strategies. We introduce a parameter called "focus length" (a positive number such as 5 or 10) to control the speed and sensitivity for our multi-AI to adapt to the opponent's strategy change. The focus length is the number of previous rounds that the multi-AI should look at when determining which Single-AI has the best performance and should choose to play for the next game. We experimented with 52 different people, each playing 300 rounds continuously against one specific multi-AI model, and demonstrated that our strategy could win against more than 95% of human opponents.

The Rock-Paper-Scissors (RPS) game has been widely used to study competitive phenomena in society and biology, such as ecological interactions, the maintenance of biodiversity in ecological systems[1–9] and price dispersion of markets[10,11]. There are two general approaches for RPS play, namely, Bayesian equilibrium and exploitation of player pattern[12,13]. The payoff parameter $a$, defined as the incentive for winning divided by the incentive for drawing, is set to 2 to form a neutral RPS game[14].

Previous research has found that there is a social circle in human competitive strategy when playing iterated RPS games[12]. In this article we proposed a multi-AI algorithm that can exploit human strategy and win against human players in the same iterated RPS games and we conduct experiments with human players to confirm the results.

Our work may stimulate future more refined experimental and theoretical studies on the microscopic mechanisms of decision-making and learning in basic game systems[15–19].

Markov chain models[20] are the single models which our multi-AI is composed of. A Markov chain is a stochastic model describing a sequence of possible events in which the probability of each event depends only on the state attained in the previous events. Here, Markov chain is a special sort of belief network used to represent the sequences of states in a dynamic system. Previous research has used Markov process model to describe the stochastic evolution dynamic of the Rock–Scissors–Paper Game[21]. Here the iterated RPS game is considered as a Markov process and Markov chains are built throughout the process of 300 rounds of competition. The models are built to exploit attempted circular exploitation patterns.

The simplest discrete-time Markov chain is the first-order Markov chain, where the probability of moving to the next state depends only on the present state and not on the previous states:

$$\Pr(X_{n+1} = x | X_1 = x_1, X_2 = x_2, \dots, X_n = x_n) = \Pr(X_{n+1} = x | X_n = x_n),$$

where $X_1, X_2, \dots X_n$ are a sequence of random variables ("Rock", "Paper", and "Scissors" here). What you will play in the next round only depends on what you played this round, like a short memory pattern sequence.

[1]National Engineering Research Center for Optical Instruments, Centre for Optical and Electromagnetic Research, Zhejiang University, Hangzhou 310058, China. [2]Ningbo Research Institute, Zhejiang University, Ningbo 315100, China. [3]Department of Electromagnetic Engineering, School of Electrical Engineering, Royal Institute of Technology, 100 44 Stockholm, Sweden. ✉email: Sailing@kth.se

Focus length F = 5

| Player played: | R | S | P | S | S | P | R | R | P | S | R | S | S | P | S | P | S | R | P | S |
|---|---|---|---|---|---|---|---|---|---|---|---|---|---|---|---|---|---|---|---|---|
| AI 1 played: | S | S | S | R | R | R | R | R | P | P | R | R | P | S | R | P | R | S | R | R |
| AI 1 Score | 1 | -1 | 1 | -1 | 0 | 1 | 0 | 0 | -1 | 1 | 0 | 0 | 1 | 1 | -1 | -1 | -1 | 1 | -1 | 1 |
| AI 1 Last 5 Rounds Total Score: | 0 | 1 | 0 | 1 | 0 | 0 | 0 | 1 | -1 | 0 | 0 | 0 | 1 | 3 | 2 | 3 | 2 | 0 | -2 | -3 | -5 |
| AI 2 played: | P | P | S | P | P | S | R | S | R | S | R | S | R | S | P | R | R | S | P | R |
| AI 2 Score | 1 | -1 | 1 | 0 | -1 | 1 | 0 | -1 | 1 | 1 | 0 | 0 | 0 | -1 | 1 | -1 | -1 | 1 | 1 | 1 |
| AI 2 Last 5 Rounds Total Score: | 0 | 1 | 0 | 0 | 1 | 1 | 0 | 0 | 1 | -1 | -2 | -1 | -2 | -1 | 0 | 1 | 0 | 1 | -1 | 0 | -1 | -1 |
| AI 3 played: | S | P | P | P | S | P | P | P | R | R | R | S | P | R | P | R | P | R | S | P |
| AI 3 Score | -1 | -1 | 0 | -1 | 0 | 0 | 1 | 1 | -1 | 1 | 0 | 1 | 0 | -1 | -1 | 1 | -1 | 0 | 1 | 0 |
| AI 3 Last 5 Rounds Total Score: | 0 | -1 | -2 | -2 | -3 | -3 | -2 | 0 | 1 | 1 | 2 | 2 | 2 | 1 | 1 | -1 | -2 | -4 | -5 | -4 | -4 | -3 |
| AI 4 played: | R | S | S | P | P | R | R | S | R | S | R | S | R | P | S | R | P | S | S | S |
| AI 4 Score | 0 | 0 | 0 | 1 | -1 | -1 | 0 | -1 | 1 | -1 | 0 | 0 | 0 | 1 | 1 | 1 | 0 | 0 | 0 | 0 |
| AI 4 Last 5 Rounds Total Score: | 0 | 0 | 0 | 1 | 0 | -1 | -2 | -2 | -4 | -4 | -2 | -2 | -1 | -1 | 1 | 1 | 3 | 2 | 3 | 2 | 0 | -1 |
| AI 5 played: | P | P | S | P | P | R | R | S | P | R | R | R | R | P | P | R | S | R | S | R |
| AI 5 Score | 1 | -1 | 1 | -1 | -1 | 0 | 0 | 0 | 1 | -1 | 0 | 1 | 0 | -1 | -1 | 0 | 1 | 0 | 1 | 1 |
| AI 5 Last 5 Rounds Total Score: | 0 | 1 | 0 | 1 | 0 | 0 | -1 | 0 | -1 | 0 | -1 | 0 | 0 | -2 | 0 | 0 | 0 | 1 | 1 | 1 | -1 | 0 | -2 | -2 | -1 |
| Multi AI = | AI1 | AI1 | AI1 | AI1 | AI2 | AI1 | AI1 | AI1 | AI3 | AI3 | AI3 | AI3 | AI1 | AI1 | AI1 | AI1 | AI4 | AI4 | AI4 | AI2 |
| Multi AI played: | P | P | S | P | P | S | R | S | R | S | R | S | R | S | P | R | R | R | P | R |

Multi AI switched, since for the recent 5 rounds, best performance model has changed.

**Table 1.** The internal calculation of our multi-AI algorithm with 5 models and focus length 5 playing against a human opponent.

Markov chains can be generalized to cases of short-term dependency, by taking into account recent past states in the chain. The m-th order Markov chain[22] considers the current state to depend on m previous states, where m is finite, and is a process satisfying

$$\Pr(X_n = x_n | X_{n-1} = x_{n-1}, X_{n-2} = x_{n-2}, \ldots, X_1 = x_1)$$
$$= \Pr(X_n = x_n | X_{n-1} = x_{n-1}, X_{n-2} = x_{n-2}, \ldots, X_{n-m} = x_{n-m}) \text{ for } n > m$$

Here the m-th order Markov chain is like a model with memory length m, which 'remembers' the previous m states.

For $n < m$, all m-th order single Markov chain models will select Rock, Paper, Scissors randomly, with 1/3 probability each.

As the competition goes, the combined model (multi-AI) will select one specific fixed memory length Markov Model that is better at predicting this particular human player's decision strategy at this particular time.

Here for simplicity we use AI-m to denote our single Markov chain model of order m.

Through the experiments we found that different models work best against different human opponent's competition strategies and the prediction results vary greatly so we built the 1st -5th order Markov chain models (i.e. AI-1 to AI-5) with different memory lengths for exploiting different human competition strategies. To make a multi-AI model that can differentiate and adapt to different human opponents, we combine the 1st-5th order single Markov models and introduce a "focus length" parameter to control the adaptation speed and sensitivity to form a multi-AI that can adapt to different human strategies and win against most of its opponents. The focus length, F, is the number of rounds that are used to evaluate the current performance of the single AIs. The multi-AI will produce the output associated with the single-AI that has the highest score in the last F rounds. Table 1 illustrates how this multi-AI model competes against a specific player as an example with focus length F = 5.

For example, for the fifth round, the multi-AI will look at all the previous 4 rounds and calculate each single models' scores. Since AI 2 had the highest score, the dominant AI will be AI 2 and its output is used as the fifth round multi-AI output. For the 9th round, the multi-AI will look at the 4th–8th rounds (focus length F = 5), for which AI 3 has the highest total score, and the dominant AI will be switched to AI 3 and its output is used as the 9th round multi-AI output.

For the first round, the multi-AI will use the result from AI-1 and it selects Rock, Paper or Scissors randomly with 1/3 probability each.

Focus length parameter F is set to control the speed and sensitivity for our multi-AI model to adapt to the opponent's strategy change. Our multi-AI model will look at the recent F rounds of history to decide which single model is currently performing the best and should produce the next output. For the first 4 rounds when our competition data is less than focus length F, which is set to 5 as before, multi-AI will simply consider all rounds to determine its next round's dominant single AI model. In the specific case of Table 1, the dominant AI for all of the first 4 rounds is AI-1. In round 5, AI-2 has the best cumulative score and thus is the dominant AI.

Table 1 shows how our multi-AI algorithm competes against a specific player when F = 5 as an example. The transition matrix for the last AI-2 seeing the player played "PS" in the past 2 rounds is:

| | R | P | S |
|---|---|---|---|
| PS | 1/3 | 1/3 | 1/3 |

| | | | | | | | | | Next Round AI |
|---|---|---|---|---|---|---|---|---|---|
| AI 1 | W | W | D | W | W | D | L | L | AI 4 |
| AI 4 | L | L | L | W | D | W | W | D | |

**Table 2.** Selection between AI-1 and AI-4 when focus length F = 5.

| | RR | RS | RP | PR | PP | PS | SR | SP | SS |
|---|---|---|---|---|---|---|---|---|---|
| R | 0 | 1/18 | 1/18 | 0 | 0 | 1/18 | 1/18 | 3/18 | 1/18 |
| P | 0 | 0 | 0 | 1/18 | 0 | 2/18 | 0 | 1/18 | 0 |
| S | 1/18 | 1/18 | 0 | 0 | 0 | 1/18 | 1/18 | 0 | 2/18 |

**Table 3.** The transition matrix for AI-2 after 20 rounds of competition.

Thus, for the next round, AI-2 has 1/3 probability to select Paper, Scissors and Rock.

Table 2 shows an example for the selection between AI-1 and AI-4 when focus length F = 5. Although globally AI-1 has a higher score in total (from the first round AI-1 has a total score of 2, but its recent 5 rounds score is 0), AI-4 has a higher score (which is 3 in this case) locally during the recent 5 rounds. Thus, the next round multi-AI will pick AI 4's result as our multi-AI output.

The transition matrix for AI-2 after 20 rounds of competition is in Table 3.

## Results

All experiments are conducted with money incentive. We did the experiment with 52 human subjects recruited at Zhejiang University and used a multi-AI model which has 5 or 10 single length Markov chain models. Figure 1 shows 4 typical results of our multi-AI strategy with a combination of the 1st–5th order Markov chain models (here focus length F is also set to 5, but it can be any other integer) competing against 4 typical players for 300 rounds.

The total score for the multi-AI against 52 people are shown in Fig. 2. Against one player, multi-AI with a combination of 5 models had a score of 151 (i.e., 198 wins, 55 draws and only 47 losses). The total times of wins for multi-AI is more than 4 times that for this human player. Figure 1d is an example of a player who beat the AI by a margin of 4, and this is not a statistically meaningful margin as there are several lead changes.

52 people's preferences in choosing Rock Paper Scissors in their 300 rounds competition are shown in Table 4. There is a slight preference in choosing Rock (Table 5).

41 players played against the multi-AI with focus length F = 5 and AI-1 to AI-5(see blue bars in Fig. 2) and 11 players played against the multi-AI with focus length F = 10 and AI-1 to AI-10 (see orange bars in Fig. 2). From the overall results in Table 5, we see that our multi-AI algorithm with F = 10 give similar scores, but has lower standard deviation than that with F = 5.

For simplicity, we let multi-5AI denote our multi-AI model with a combination of the 1st–5th order Markov chain models (here focus length F is also set to 5, but it can be any other integer), and multi-10AI denote our multi-AI model with a combination of the 1st–10th order Markov chain models (focus length F is also set to 10).

We looked at the performance of individual models within a multi-AI 300 game set. The AI which performs the best against a particular individual varies greatly, but overall AI 2–6 perform better than AI 1 or higher-order Markov chains of longer memory length. This general trend is consistent with human short memory holding around 7 items[23].

Figures 3 and 4 show AI with different memory lengths' performance against specific human players.

It is hard to build one single model that can exploit every different human's behavior, and thus we decided to combine the single models to make it able to differentiate and adapt to more different human competition strategies and win against most of its opponents.

## Discussion and conclusions

In this paper, we have introduced a multi-AI model that wins against human in iterated RPS games and experimentally confirmed our results. We found that using a single-length Markov model could beat most human players, but not all human players.

In single model experiments, we found the model with the best performance varies greatly for different people, which indicates that different people have different patterns. Although different humans have different patterns, and in total the patterns may be very hard to observe and exploit. Human competition behavior indeed has patterns and the patterns are exploitable by using proper simple models (single models that successfully predict this human's behavior). We have obtained and exploited different human behaviors by building and combining single

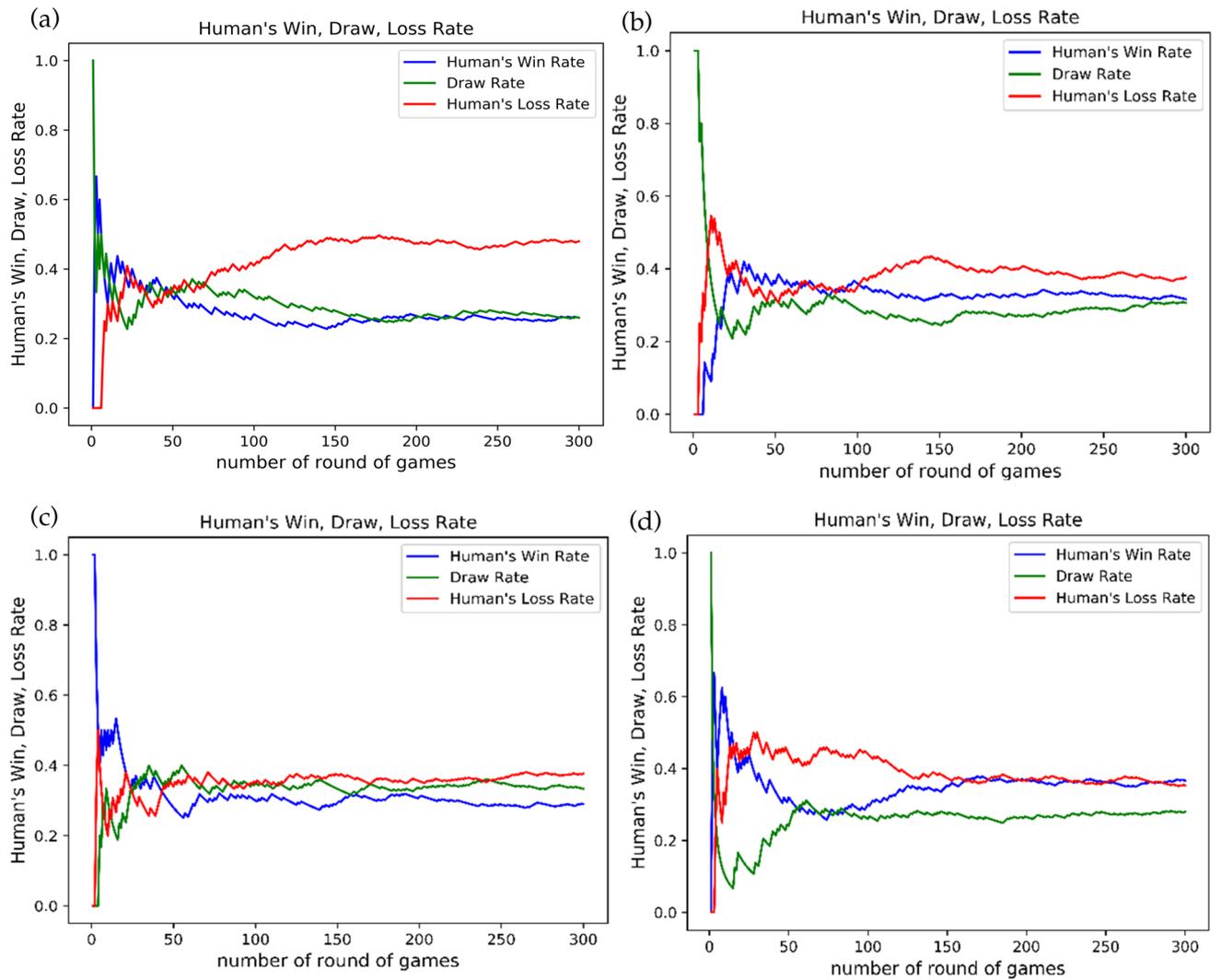

**Figure 1.** 300 rounds AI competition results for 4 typical players.

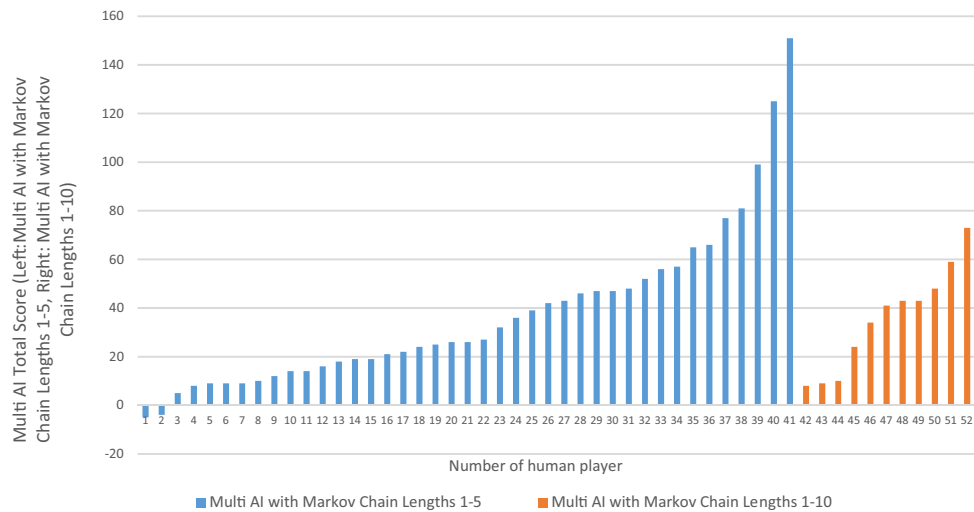

**Figure 2.** Total scores for multi-AI competing against different players in 300 rounds game.

| | R | P | S |
|---|---|---|---|
| MEAN (times) | 106.6098 | 96.29268 | 97.09756 |
| STDEV.S | 13.08095 | 13.12239 | 12.11851 |

**Table 4.** 52 people's preferences in choosing Rock Paper Scissors in 300 rounds competitions.

| | AI1 | AI2 | AI3 | AI4 | AI5 | AI6 | AI7 | AI8 | AI9 | AI10 | 10 single models average score | Multi-10AI score | Multi-5AI score |
|---|---|---|---|---|---|---|---|---|---|---|---|---|---|
| Player1 | −13 | 37 | 24 | 20 | 22 | 14 | −6 | −2 | −10 | 3 | 8.9 | 10 | |
| Player2 | 2 | 50 | 48 | 42 | 42 | 50 | 61 | 7 | 12 | 5 | 31.9 | 41 | |
| Player3 | 46 | 58 | 52 | 53 | 24 | 29 | 11 | −15 | −14 | 12 | 25.6 | 8 | |
| Player4 | 26 | 41 | 26 | 55 | 46 | 28 | 20 | 13 | 10 | 36 | 30.1 | 34 | |
| Player5 | 35 | 39 | 92 | 107 | 85 | 67 | 88 | 61 | 78 | 70 | 72.2 | 73 | |
| Player6 | 28 | 60 | 52 | 55 | 49 | 36 | 14 | 1 | −18 | 12 | 28.9 | 43 | |
| Player7 | 56 | 63 | 68 | 65 | 55 | 46 | 54 | −1 | −14 | 17 | 40.9 | 48 | |
| Player8 | 28 | 29 | 47 | 29 | 12 | 5 | 21 | 5 | −10 | −5 | 16.1 | 43 | |
| Player9 | 32 | 63 | 58 | 72 | 62 | 72 | 32 | 26 | 20 | 15 | 45.2 | 59 | |
| Player10 | −5 | −14 | 8 | 11 | 16 | −7 | −26 | 19 | 20 | 23 | 2.9 | 24 | |
| Player11 | 40 | 15 | 33 | 16 | 9 | 24 | 25 | −9 | 6 | 37 | 19.6 | 9 | |
| MEAN | 25.00 | 40.09 | 44.73 | 47.73 | 38.36 | 33.09 | 26.73 | 9.55 | 7.27 | 20.45 | 29.30 | 35.64 | 37.39 |
| STDEVA | 21.61 | 23.57 | 26.02 | 28.36 | 23.93 | 24.52 | 31.69 | 20.76 | 27.38 | 20.86 | 19.05 | 21.21 | 33.12 |

**Table 5.** Game results (total scores) of our multi-10AI competing with human in 300 rounds.

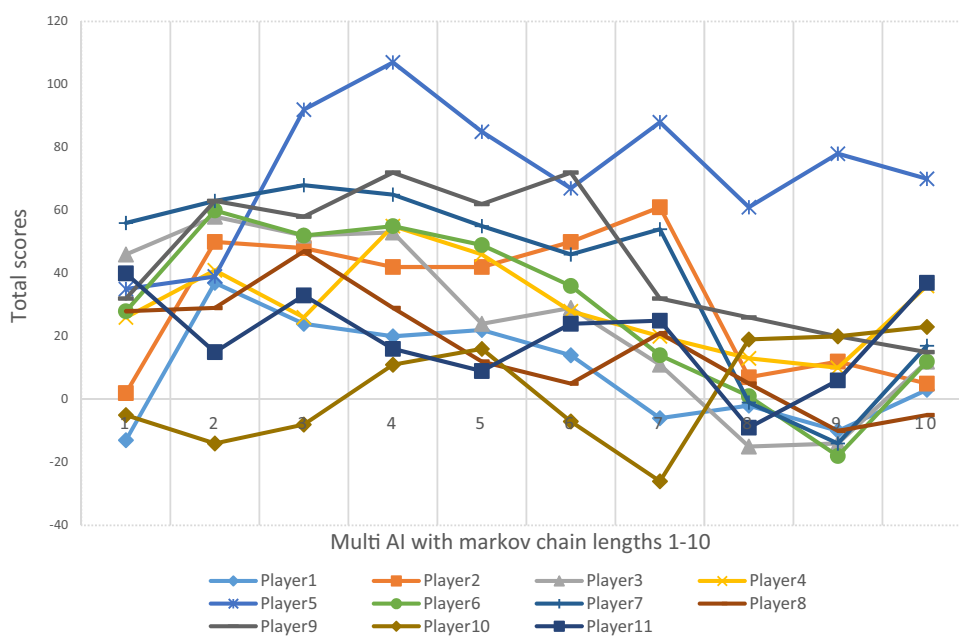

**Figure 3.** Game results (total scores) of Multi-10AI competing with human in 300 rounds.

Markov chain models of different memory lengths and during the competition process it learns and switches to the best prediction model according to its focus length. We have introduced one possible architecture for human AI RPS games competition, and this model could be further improved by e.g., optimizing the voting weights of single Markov chain models, using the first part of the competition data to pre-train multi-AI model and switch to only two or three dominant single models after the pre-training process. Focus length is a hyper parameter and can be tested by more human experiments for further optimization. After rearranging single models and adjusting "focus length" our model can potentially be improved further. The competition behavior patterns and their successful exploitation may lead our future work to better modeling, predicting and adapting to different specific human's competition behavior patterns.

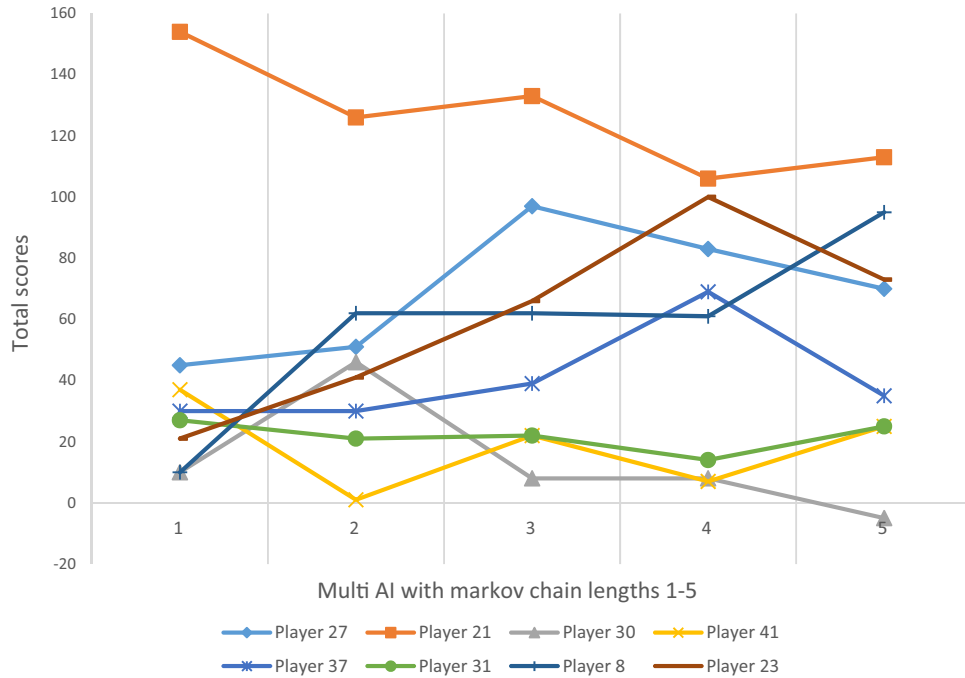

**Figure 4.** Game results (total scores) of multi-AI with Markov chain lengths 1–5 competing with 8 typical players in 300 rounds.

## Methods

**Experiment.** Our methods and regulations for this experiment mostly follow the RPS social experiments conducted in the same university by Zhijian Wang et al.[12]. The first author confirms that all methods and experiments were carried out in accordance with the guidelines and regulations of "5 Ethical Considerations in Sociological Research" in Code of Ethics made by The American Sociological Association (ASA)[24] and approved by the Ethics Committee of our center of Zhejiang University (China). The humans competing against our multi-AI in iterated RPS game experiment was approved by the relevant laboratory of our center of Zhejiang University and performed at Zhejiang University in the period of July 2019 to September 2019. A total number of 52 undergraduate and graduate students of Zhejiang University volunteered to serve as the human subjects of this experiment. These students were openly recruited at the university library and informed consent was collected from all the participating human subjects. The 52 human subjects (referred to as players in this paper) carried one experimental session by playing the RPS game for 300 rounds with fixed payoff parameter a = 2. To prepare for the experiments, the players were led and seated separately in a lab room, each facing a computer screen. They were not allowed to communicate with each other throughout the experiment. Written instructions were handed out to each player and the rules of the experiment were also orally explained by an experimental instructor. The rules of the experimental session are as follows:

   i. Each player plays the RPS game repeatedly with our computer program.
   ii. Each player earns virtual points during the experimental session where each winning round will earn the player 2 virtual points and each drawing round will earn the player 1 virtual point, and the number of virtual points won't change if the player loses. These virtual points are then exchanged into money in RMB as a reward to the player, plus an additional 5 RMB as show-up fee.
   iii. After the decision has been made it cannot be changed.

Before the formal experimental session begins, the players were asked four questions to ensure that they completely understand the rules of the experimental session. The four questions are: (1) If your opponent chooses "Scissors" and you choose "Rock", how much money will you earn for this round? (2) If your opponent chooses "Rock" and you choose "Rock", how much money will you earn for this round? (3) If your opponent chooses "Rock" and you choose "Scissors", how much money will you earn for this round? (4) Do you know that at each game round you will play with an AI opponent? During the experimental session, an information window and a decision window are displayed on the computer screen in front of each player. The window on the left of the computer screen is the information window. The current game round, the time limit (40 s) of making a choice, and the time left to make a choice are displayed on the upper panel of the information window. At the beginning of each game round, the color of this upper panel turns to green. If the player does not make a decision within 20 s, the color will change to yellow. The color will change to red if the decision time runs out (then the experimental instructor will urge the player to make a decision immediately). Fortunately, all the players made their decisions within 20 s. The color will change to blue if a decision has been made by the player. After the player has made his/her decision for this round, the lower panel of the information window will immediately show the opponent's choice, the player's choice, and the player's payoff for this game round. Each player's cumulated payoff is also shown on the screen. The players are asked to write down their own choices of each game round on the record sheet (Rock as $R$, Paper as $P$, and Scissors as $S$). The decision window is on the right side of the computer screen. The upper panel of the decision window lists the current game round, and the bottom panel lists the three candidate actions "Rock", "Scissors", "Paper" horizontally from left to right. The player can make a decision by clicking on the corresponding action word. After the decision has been made, the player will know the result for this game round and the decision window will be asking for the next game input. The reward money in RMB for each player is determined by the following formula. Suppose a player $i$ earns $x_i$ virtual points in the whole experimental session, the total reward $y_i$ in RMB for this player is given by $y_i = x_i \times r + 5$, where $r$ is the exchange rate between virtual point and money in RMB. According to the mixed-strategy Nash equilibrium, each player's expected payoff in one game round is $(1+a)/3$. Therefore, we set the exchange rate to be $r = 0.45/(1+a)$ to ensure that, under the mixed-strategy NE assumption, the expected total earning for a player will be 50 RMB regardless of the particular experimental session. The payoff parameter $a$ is set to 2. The numerical value of $r$ is 0.15. The above-mentioned reward formula and how much virtual points you will earn if you win, draw or lose were listed in the printed instruction and also verbally mentioned at the instruction phase of the experiment by the experimental instructor.

## Data availability
All 52 players experiment data with all process data of Markov chains we used are provided in the Supplementary Informations 1 and 2.

## Acknowledgements

This work was partially supported by the National Key Research and Development Program of China (No. 2018YFC1407506), the National Natural Science Foundation of China (No. 11621101), and the Fundamental Research Funds for the Central Universities (Zhejiang University NGICS Platform). We thank Shuo Li for excellent assistance in arranging the experiments. Open access funding provided by Royal Institute of Technology.

## Author contributions

S.H. conceived and supervised the work. L.W. developed the algorithm and built the multi-AI. L.W., Y. L. and W.H. performed the experiments and collected data. L.W. processed and analyzed the data. The manuscript was discussed and written by L.W., J.E. and S.H. with comments and input from all authors.

## Competing interests

The authors declare no competing interests.

## Additional information

**Supplementary information** is available for this paper at https://doi.org/10.1038/s41598-020-70544-7.

**Correspondence** and requests for materials should be addressed to S.H.

**Reprints and permissions information** is available at www.nature.com/reprints.

**Publisher's note** Springer Nature remains neutral with regard to jurisdictional claims in published maps and institutional affiliations.